\documentclass[aps,prb,amsfonts,amssymb,twocolumn,amsmath,preprintnumbers,nofoot
inbib,floatfix,showpacs]{revtex4}
\usepackage{graphicx}
\usepackage{bm}
\usepackage{amsmath}
\usepackage{amssymb}

\begin{document}



\title{Effects of spin-orbit interaction on superconductor-ferromagnet
heterostructures: spontaneous electric and spin surface currents}
\author{I.~V.~Bobkova}
\affiliation{Institute of Solid State Physics, Chernogolovka,
Moscow reg. 142432, Russia}
\author{Yu.~S.~Barash}
\affiliation{Institute of Solid State Physics, Chernogolovka,
Moscow reg. 142432, Russia}



\begin{abstract}
We find proximity-induced spontaneous spin and electric
surface currents, at all temperatures below the superconducting
$T_c$, in an isotropic $s$-wave superconductor deposited with a thin
ferromagnetic metal layer with spin-orbit interaction. The currents
are carried by Andreev surface states and generated as a joint effect
of the spin-orbit interaction and the exchange field. The background
spin current arises in the thin layer due to different local spin
polarizations of electrons and holes, which have almost opposite
velocities in each of the surface states. The spontaneous surface
electric current in the superconductor originates in asymmetry of
Andreev states with respect to sign reversal of the momentum
component parallel to the surface.  Conditions for electric and spin
currents to show up in the system, significantly differ from each
other.
\end{abstract}





\maketitle

Proximity effects in superconductor-ferromagnet heterostructures have
attracted much attention for recent years. In contrast with the
nonmagnetic case, magnetic surfaces and interfaces make spin-flip
processes possible and suppress an $s$-wave superconducting order
parameter, generating Andreev bound states in adjacent superconducting
regions \cite{fogel00,fogel01,nazarov01,bb02,bbk02}. Spin structure of
Andreev bound states near complex magnetic interfaces can be rather
involved \cite{bbk02}. Triplet components of the order parameter in a
singlet superconductor can be induced by ferromagnets under certain
conditions \cite{bve03}. Cooper pair wave functions exponentially
decay into the bulk of ferromagnets, oscillating at the same time
\cite{bbp82}, and acquire a triplet component in the ferromagnetic
region \cite{bve01}. These proximity effects can lead, in particular,
to specific properties of the Josephson current through magnetic
interfaces, which have been intensively studied both theoretically
and experimentally \cite{bbp82,mrs88,buz92,demler97,fogel00,fogel01,%
nazarov01,volkov01,ryazanov01,aprili02,bb02,bbk02,bve03}.
Also, proximity-induced nonmonotonic dependence of the superconducting
critical temperature on the thickness of the ferromagnetic layer has
been thoroughly studied for superconductor-ferromagnetic metal
bilayers or heterostructures (see, for example,
\cite{buz92,demler97,fominov,volkov02,buzdin04,bader04} and references
therein).

Spontaneous surface currents represent other important example of
possible proximity-induced effects. Spontaneous electric currents,
taking place near surfaces or interfaces on the scale of the
superconducting coherence length, produce a magnetic field and,
hence, a counterflow of screening supercurrents on the
scale of the penetration depth. The electric current can arise, for
example, near nonmagnetic surfaces and interfaces of unconventional
superconductors, whose states break time-reversal symmetry
\cite{volgor85,sigueda91}. In particular, the electric current,
carried by Andreev states, appears near nonmagnetic surfaces and
interfaces of chiral superconductors \cite{vol97}. The
current also occurs, if a surface-induced subdominating pairing shows
up near surfaces of $d$-wave superconductors, breaking time-reversal
symmetry of the superconducting state \cite{frs97}. Other possible
mechanism generating electric surface currents, is specifically based
on a paramagnetic response of the zero-energy Andreev surface states
to an applied magnetic field. This can take place at low temperatures
at smooth $(110)$ surfaces of $d$-wave superconductors
\cite{bkk00,wendin01}, as well as in a system with a thin
ferromagnetic metallic layer deposited on a semi-infinite bulk
isotropic $s$-wave superconductor \cite{annett02}. In the latter case
the energy of the surface states becomes zero only for several values
of the layer thickness and, in the presence of particle-hole asymmetry,
the spontaneous electric current is accompanied by a spontaneous
surface spin current \cite{annett02}. Dissipationless background spin
currents, which take place in various systems in the equilibrium and
do not lead to any spin accumulation, have been a subject of recent
discussions and studies \cite{macdonald01,rashba03,pareek04}. The
spin currents can be generated, for example, by the spin-orbit
interaction (in particular, via Rashba term) in two-dimensional
metals. Measurements of background persistent spin currents are not
carried out for now, although some suggestions for a direct detection
of these currents have been proposed in the literature
\cite{macdonald01,loss03}.

In the present paper we study spontaneous currents under conditions,
when spin-orbit interaction takes place in a thin ferromagnetic metal
layer in proximity to an isotropic $s$-wave superconductor. Joint
effect of the spin-orbit interaction, described by the Rashba term, and
the exchange field is shown to play an important role in generating
spontaneous currents. We find that superconductor induces background
spin currents in the ferromagnetic layer with the spin-orbit
interaction (FSOL) for all temperatures below the superconducting
$T_c$. This spin current is carried by Andreev surface states and
takes finite values due to different local spin polarizations of
electrons and holes, which have almost opposite velocities in each of
these states. Maximal possible values of the background spin current
density are the order of the Landau depairing current
density. We find proximity-induced finite spin currents within the
quasiclassical approach, when only linear terms in small parameters
$\alpha p_f/\varepsilon_f$, $h/\varepsilon_f$ are taken into account in
describing the FSOL. Spontaneous background spin currents, arising in the
two-dimensional electron systems with spin-orbit interaction without
any proximity effects \cite{rashba03}, contain higher powers of these
parameters which are assumed small below. Further, a spontaneous
electric surface current, carried by Andreev surface states, arises
in the superconductor due to proximity to the FSOL. Respective
structures of wave functions and spectra of the surface states are
strongly influenced by the spin-orbit interaction and the exchange
field and differ for quasiparticles with opposite momentum components
${\bm p}_{f\|}$ parallel to the surface. The spontaneous electric
current arises as a result of this asymmetry of Andreev states with
respect to ${\bm p}_{f\|}\to-{\bm p}_{f\|}$, to some extent analogously
to the current induced by chiral surface or interface states.
Conditions for electric and spin currents to show up in the system we
study, significantly differ from each other. Thus, the spontaneous
spin current in the FSOL arises even within the framework, when the
surface electric current vanishes.

Consider an isotropic $s$-wave superconductor at $x>d$, deposited
with a layer of thickness $d$ made of a ferromagnetic metal. Let a
macroscopic thickness of the layer be much less than the
superconducting coherence length: $d\ll \xi_s$. Both the internal
exchange field $\bm h$ and spin-orbit Rashba term
$\bm{w}\bm{\sigma}=\alpha({\bm n}\times {\bm
p}_{||})\bm{\sigma}$ enter the Hamiltonian density of the FSOL:\,
$\hat{\cal H}(x)=\hat{H}^{(0)}-\left(\bm{h}(x)+\bm{w}(x)
\right)\bm{\sigma}$. Here $\hat{H}^{(0)}$ describes the kinetic
energy of free electrons, $\bm{n}$ is the unit vector along the
surface normal and $\bm{p}_{||}$ the momentum component parallel
to the surface. The exchange field is assumed always aligned along
the $z$-axis. Both $\bm{h}(x)$ and $\bm{w}(x)$ are taken finite
and spatially constant within the FSOL\, $0<x<d$. The $x$-axis is
taken directed into the depth of the superconductor and the system
is confined by an impenetrable wall at $x=0$.

We assume $\Delta\ll h, \alpha p_{f}\ll \varepsilon_f$ and describe
the system in question by quasiclassical Eilenberger equations
for Matsubara Green's function:
\begin{equation}
-i v_{f,x} \dfrac{\partial \check g}{\partial x} = \left[ \left( i
\varepsilon_n \hat \tau_z +\hat \tau_z \check \Delta + \bm h
\check {\bm s}+ \bm w \hat \tau_z \check {\bm s} \right),
\check g \right] \label{eilenberger} \enspace ,
\end{equation}
\begin{equation}
\check g^2 = - \pi^2 \label{norm} \enspace .
\end{equation}
Here $\check{g}(x,{\bm p}_f,\varepsilon_n)$ takes $4\times4$ matrix form in
the four-dimensional product space of particle-hole and spin variables. In the
particle-hole space
\begin{equation}
\check g(\bm p, \varepsilon_n, x) = \left(
\begin{array}{cc}
{\hat g}(\bm p, \varepsilon_n, x) & \hat f(\bm p, \varepsilon_n, x) \\
\hat {\widetilde f~}(\bm p, \varepsilon_n, x) &
\hat {\widetilde g}(\bm p, \varepsilon_n, x)
\end{array}
\right) \label{structg} \enspace ,
\end{equation}
where all matrix elements are $2\times2$ matrices in spin space.
Pauli-matrices in particle-hole space are $\hat\tau_j$, $\hat
\tau_{\pm} = \hat \tau_x \pm i \hat \tau_y$, while in spin space
$\hat \sigma_i$. The superconducting order parameter matrix is
$\check \Delta = 1/2 [\hat \tau_+\Delta - \hat \tau_-
\Delta^*]i\hat\sigma_y$. The operator for quasiparticle spin is
$({1}/{2})\check{\bm s}\hat\tau_z$, whereas the operator $\check
{\bm s} =1/2[ (1+\hat\tau_z)\hat{\bm \sigma} - (1-
\hat\tau_z)\hat \sigma_y \hat{\bm \sigma} \hat \sigma_y]$ enters the
Zeeman term. The order parameter $\Delta$ is taken spatially constant
throughout the superconducting half-space $x>d$.  As this follows
from recent results for two-dimensional superconductors with
spin-orbit coupling \cite{feigel03}, a possibility for
proximity-induced inhomogeneous phase of the order parameter in the
plane parallel to the interface should be studied for sufficiently
thin superconducting layer in proximity to the FSOL. This
two-dimensional inhomogeneous profile of the phase does not appear,
however, for a massive superconducting sample.

Electric and spin current densities can be expressed via
quasiclassical Green's function as follows
\begin{eqnarray}
\bm j = N_f T\bigl< \bm v_{f} \sum \limits_{\varepsilon_n}{\rm
Sp}_2\hat g({\bm p}_f,\varepsilon_n)\bigr>_{S_f},
\label{elcur} \\
\bm j_{i}^s=\dfrac{ N_f T}{ 2}\bigl< \bm v_{f}
\sum \limits_{\varepsilon_n}{\rm Sp}_2\hat
\sigma_i\hat g({\bm p}_f,\varepsilon_n)\bigr>_{S_f} .
\label{spincur}
\end{eqnarray}
Here $N_f$ is the normal state density of states per spin direction,
\mbox{$<\ldots >_{S_f}$} means averaging over quasiparticle states at
the Fermi surface. Spin current $j^s_{il}$ carries i-th
spin component along $l$-axis in coordinate space. One can introduce
scalar $g_0$ and vector $\bm g$ components of the Green's function in
spin space $\hat g({\bm p}_f,\varepsilon_n)=g_0({\bm p}_f,\varepsilon_n){
\hat\sigma}_0+{\bm g}({\bm p}_f,\varepsilon_n)\hat{\bm \sigma}$.
As this is seen from Eqs.(\ref{elcur}), (\ref{spincur}), $g_0$ determines
electric current, while spin current is associated with ${\bm g}$.

The Green's function for the FSOL satisfies conventional boundary
conditions on the impenetrable wall at $x=0$:
$\check{g}(0,{\bm p}_f,\varepsilon_n)=\check{g}(0,\tilde{\bm p}_f,
\varepsilon_n)$, where ${\bm p}_f$ and $\tilde{\bm p}_f$ are the incoming
and the outgoing quasiparticle momenta respectively. We match solutions
of Eilenberger equations for the superconduting half-space and for the FSOL
with the continuity conditions on a transparent interface at $x=d$.
Substituting the final result for the Green's function into Eqs.(\ref{elcur}),
(\ref{spincur}), we find no spontaneous electric current in the system and
finite components $j^s_{yz}$, $j^s_{zy}$ of spin current situated in the FSOL
and flowing parallel to the surface.

One can show that the whole spin current is carried by Andreev
surface states taking place in the system. We find two dispersive branches
of Andreev surface states, whose energies depend on momentum component
parallel to the surface:
\begin{equation}
\varepsilon_{1,2} = \mp{\rm sgn}\left[\sin
\left(\dfrac{\Phi}{2}
\right)\right] \Delta\cos \left( \dfrac{\Phi}{2} \right)
\enspace .
\label{bsFSO}
\end{equation}
Here
\begin{eqnarray}
\cos\Phi=\cos^2 \dfrac{\varphi}{2}
\cos\dfrac{\Theta_+ \!+\! \Theta_-}{2} + \sin^2 \dfrac{\varphi}{2}
\cos \dfrac{\Theta_+ \!-\! \Theta_-}{2}, \label{Phi}\\
\nonumber\\
\Theta_{\pm}=\dfrac{4 |\bm h \pm \bm w|d}{|v_{f,x}|},
\quad \cos\varphi={\bm e}_+{\bm e}_-,
\quad {\bm e}_{\pm}=\dfrac{(\bm h\pm \bm w)}{ |\bm h\pm \bm w|}
\label{Theta} .
\end{eqnarray}
In the absence of spin-orbit interaction spectra of Andreev states, descibed
by Eqs. (\ref{bsFSO}), (\ref{Phi}), (\ref{Theta}), reduce to the results for
spin-discriminated Andreev states at a ferromagnetic surface
\cite{fogel00,bb02}.

Andreev surface states carry no spin current in a singlet
superconductor, since particles and holes, occupying the state, have
identical spatially constant local spin polarization and opposite
velocities. However, the wave function of Andreev surface states does
not vanish in the FSOL and has a qualitatively different spin
structure there, as compared with the superconducting region. One can
extract pole-like terms from the whole expression for the electron
retarded Green's function $\hat{g}^R(x,{ \bm p}_f,\varepsilon )$ near
bound state energies $\varepsilon_{1,2} $. We determine the spin
structure of electrons in the Andreev states in terms of eigenvectors
of these pole-like terms in spin space ${\alpha \choose \beta}(x,
{\bm p}_f,\varepsilon)$. The unit vector ${\bm P}^e$, describing
electron spin polarization, can be found from the equation ${\bm
P}^e{\widehat{\bm \sigma}}{\alpha \choose \beta}( x, {\bm p}_f,
\varepsilon)={\alpha \choose \beta}(x, {\bm p}_f,\varepsilon)$.
As a result, we obtain the following spatially dependent spin
polarization for electrons in Andreev surface states (\ref{bsFSO})
at $0<x<d$:
\begin{eqnarray}
&&{\bm P}^e(\bm p_f, \varepsilon_{1,2})=\mp\dfrac{1}{ \sin\Phi}\left[
\left({\bm e}_+\times{\bm e}_-\right)\sin\dfrac{\Theta_+ x}{ 2d}
\sin\dfrac{\Theta_-}{ 2}-\right. \nonumber\\
&&\left.
-\bigl({\bm e}_+\times\bigl({\bm e}_+\times{\bm e}_-
\bigr)\bigr)\sin\dfrac{\Theta_-}{2}\left(\cos\dfrac{\Theta_+ x}{2d} -
\cos \dfrac{\Theta_+}{2}\right)\!+\nonumber \right.\\
&&\left.
+\left({\bm e}_-\cos \dfrac{\Theta_+}{2}
\sin \dfrac{\Theta_-}{2}+{\bm e}_+\cos \dfrac{\Theta_-}{2}\sin\dfrac{\Theta_+}{2}
\right)\right].
\label{P}
\end{eqnarray}
Local spin polarization of electrons, occupying Andreev states, is
spatially constant inside the superconductor and takes there the same
value as follows from Eq.(\ref{P}) at $x=d$.
Parallel and normal to the surface components of spin polarizations, taken
for incoming and outgoing electrons in one and the same Andreev state,
are related with each other as ${\bm P}_{\|}^e(\tilde{\bm
p}_f,\varepsilon_{1,2}) = {\bm P}_{\|}^e(\bm p_f,\varepsilon_{1,2})$,\,
${\bm P}_{\perp}^e(\tilde{\bm p}_f,\varepsilon_{1,2}) = -{\bm P}_{\perp}^e(
\bm p_f,\varepsilon_{1,2})$. Also, since $\varepsilon_1=-\varepsilon_2$,
we find from Eq.(\ref{P}), that ${\bm P}^e({\bm p}_f,-\varepsilon_m)=
-{\bm P}^e({\bm p}_f,\varepsilon_m)$.

Spin polarization ${\bm P}^h$ for holes, occupying Andreev states,
can be derived from Eq.(\ref{P}). The quantity ${\bm P}^h$ satisfies
the equation $-{\bm P}^hi{\widehat{\sigma}}_y{\widehat{\bm\sigma}}i{
\widehat{\sigma}}_y{\alpha_h\choose \beta_h}(x, {\bm p}_f,\varepsilon)
={\alpha_h \choose \beta_h}(x, {\bm p}_f,\varepsilon)$, which
contains the spin operator for holes $-(1/2)i{\widehat{\sigma}}_y{
\widehat{\bm\sigma}}i{\widehat{\sigma}}_y$. Here
${\alpha_h \choose \beta_h}(x, {\bm p}_f,\varepsilon)$ is the eigenvector
of the pole-like term in the Green's function $\hat{\tilde g}^R(x,{\bm p}_f,
\varepsilon)$ near $\varepsilon_1$ or $\varepsilon_2$.
As this follows from the general relation $\hat{\tilde g}^R(x,{\bm p}_f,
\varepsilon)=\hat{g}^{A^*}(x,-{\bm p}_f,-\varepsilon)$,  the eigenstates
for holes and for electrons are associated with each other as ${\alpha_h
\choose \beta_h}(x, {\bm p}_f,\varepsilon)={\alpha^* \choose \beta^*}(x,
-{\bm p}_f,-\varepsilon)$. Hence, the spin polarization for holes
in the state ${\alpha_h \choose \beta_h}(x, {\bm p}_f,\varepsilon)$
coincides with that for electrons in the state ${-\beta^*\choose \alpha^*}
(x, -{\bm p}_f,-\varepsilon)$. Further, as this follows from the equation
for ${\bm P}^e$ and the relation between  ${\bm P}^e({\bm p}_f,\varepsilon_{m}
)$ and ${\bm P}^e({\tilde{\bm p}}_f,-\varepsilon_{m})$, the quantity
${\bm P}^h({\bm p}_f,\varepsilon_m)$ coincides with electron spin polarization
in the state ${\alpha^*\choose -\beta^*}(x, p_{fx},-{\bm p}_{f\|},\varepsilon_m)
$. Comparing electron spin polarizations of states ${\alpha^*\choose -\beta^*}(
x, p_{fx},-{\bm p}_{f\|},\varepsilon_m)$ and ${\alpha\choose \beta}(
x, p_{fx},-{\bm p}_{f\|},\varepsilon_m)$, we find finally that
spin polarization for holes can be found from Eq.(\ref{P}) as ${\bm
P}_{\|}^h(\bm p_f, \varepsilon_{1,2}) = {\bm P}_{\|}^e(p_{fx},-\bm p_{f\|},
\varepsilon_{1,2})$, $ {\bm P}_{\perp}^h(\bm p_f,\varepsilon_{1,2})=
-{\bm P}_{\perp}^e(p_{fx},-\bm p_{f\|},\varepsilon_{1,2})$.

The first term in Eq.(\ref{P}) describes ${\bm P}^e_{\perp}$
component of the spin polarization, while the second and third terms
determine ${\bm P}^e_{\|}$. Under the transformation
${\bm p}_{f\|} \to -{\bm p}_{f\|}$ one finds ${\bm e}_\pm\to{\bm e}_{\mp}$
and $\Theta_\pm\to\Theta_{\mp}$. The first two terms in the square brackets
in Eq.(\ref{P}) are responsible for a spatially dependent difference between
electron and hole local spin polarizations, taking place in Andreev states
in FSOL as the joint effect of Zeeman and spin-orbit couplings. Indeed, for
vanishing $h$ or $\alpha$ vectors ${\bm e}_{\pm}$ become parallel to each
other. Then the first and the second terms in Eq.(\ref{P}) vanish, resulting
in identical spin polarizations of electrons and holes. Also, at $x=d$
this follows from Eq.(\ref{P}) ${\bm P}^e({\bm p}_f,\varepsilon_{1,2})=
{\bm P}^h({\bm p}_f,\varepsilon_{1,2})$.

Different spin polarizations and almost opposite velocities of electrons
and holes, occupying Andreev surface states Eq.(\ref{bsFSO}), result in
a net spin current in the FSOL. The local spin current density carried by
two Andreev states can be represented as $j_{i,\|}^{s}=j_{i,\|}^{s,1}+
j_{i,\|}^{s,2}$, where
\begin{equation}
j_{i,\|}^{s,m} =\dfrac{1}{ 2}\left\langle {\bm v}_{f,\|} W_{m}
\left[ P_i^e(\bm p_f,\varepsilon_m) - P_i^h(\bm p_f,\varepsilon_m)
\right] n_f(\varepsilon_m) \right\rangle_{S_f} .
\label{current_bs}
\end{equation}
Here $W_{m}=\frac{1}{2}\pi\Delta N_f\left|\sin\frac{\Phi}{2}\right|$
is the weight of the delta-peak in the local density of states, taken
in the FSOL for $m$-th Andreev state ($m=1,2$), and
$n_f(\varepsilon)$ is the Fermi distribution function for
quasiparticles. Substitution of the represented results into
Eq.(\ref{current_bs}) gives exactly the spin current density, which
follows from Eq.(\ref{spincur}) and respective solutions of the
Eilenberger equations for the Green's function.

As this follows from Eq.(\ref{current_bs}) after integration over the
Fermi surface, only parallel to the surface components $j_{y,z}^{s}(x)$
and $j_{z,y}^{s}(x)$ of the spin current remain finite. The spin current
$j_{x,\|}^{s}$, carrying along the surface perpendicular to the surface
spin component, vanishes in accordance with the relation ${\bm
P}_{\perp}^e(\tilde{\bm p}_f,\varepsilon_{1,2})=-{\bm
P}_{\perp}^e(\bm p_f,\varepsilon_{1,2})$, since the contributions
from incoming and outgoing electrons, as well as holes, cancel each
other. The proximity-induced background spin current we have found
does not lead to any spin accumulation and satisfies the continuity
equation $\sum_l\partial j^s_{il}/\partial x_l=0$. In the limit
of small Zeeman coupling $h\ll\alpha p_f$, we find the following
simple estimations for the spin current in the thin layer with
spin-orbit interaction $d\ll v_f/(\alpha p_f)$:
\begin{equation}
j^s_{\alpha\beta}=-A_{\alpha\beta}\left(\dfrac{ h}{
 \alpha p_f}\right)^2
\left(\dfrac{ d\alpha p_f}{ v_f}\right)j_{cL}\, .
\end{equation}
Here $\alpha,\beta=y,z$ and $\alpha\ne\beta$, $A_{\alpha\beta}>0$ is a
constant the order of unity, $j_{cL}=n_s\Delta/p_f$ is the Landau depairing
current density. At low temperatures $j_{cL}\sim N_fv_f\Delta$.

In the opposite limit $h\gg\alpha p_f$, when the exchange field in the FSOL
significantly exceeds spin-orbit coupling, estimations for the two components
of the spin current give different results:
\begin{eqnarray}
j^s_{yz}=B_{yz}\left(\dfrac{\alpha p_f}{ h}\right)
\left(\dfrac{ dh}{ v_f}\right)j_{cL}\, ,\\
j^s_{zy}=-B_{zy}\left(\dfrac{\alpha p_f}{ h}\right)^3
\left(\dfrac{ dh}{ v_f}\right)j_{cL}\, .
\end{eqnarray}
Here $B_{yz}$, $B_{zy}$ are constants of the order of unity.
For $h\sim\alpha p_f\sim v_f/d$\, spontaneous spin current densities
turn out to be the order of $j_{cL}$.

Background spin current density, arising without any proximity effects in the
two-dimensional metal with Rashba spin-orbit interaction \cite{rashba03},
takes the form \, $\left({\alpha p_f}/{\varepsilon_f}
\right)^3\varepsilon_fN_fv_f/6$. It is of the third-order in parameter
${\alpha p_f}/{\varepsilon_f}$, which is presumably small.
These spin currents are carried by all the occupied states
at a given temperature \cite{rashba03,pareek04}, in contrast
with the currents induced by a proximity to the superconductor.
For this reason respective reference quantity $\varepsilon_fN_fv_f$
contains a large parameter $\varepsilon/\Delta$ as compared with $j_{cL}$.

We return now to the problem of spontaneous surface electric current.
Each separate Andreev surface state, taken for given ${\bm p}_{\|}$,
carries finite surface electric current. There is no net electric
current under the conditions considered above, since electric
currents carried by Andreev surface states Eq.(\ref{bsFSO}),
(\ref{Phi}), (\ref{Theta}) with ${\bm p}_{\|}$ and $-{\bm p}_{\|}$
cancel each other. This is associated with the symmetry of scalar
component $g_0$ of the quasiclassical Green's function with respect
to the sign reversal of the momentum parallel to the surface. Spin
current takes finite values since vector component $\bm g$ of the
Green's function does not possess the symmetry. However, the
symmetry of $g_0$ turns out to be approximate, taking place only
under the conditions $\alpha p_f, h\ll\varepsilon_f$, within the
quasiclassical approximation applied to the FSOL. For this reason we
find below finite spontaneous surface electric current, assuming
$\Delta\ll\alpha p_f, h\lesssim\varepsilon_f$ and applying the
${\check S}$-matrix approach for describing the FSOL.

A surface with the FSOL is characterized by the normal-state
scattering $\check S$-matrix, contained reflection amplitudes
for quasiparticles. The $\check S$-matrix can be represented as ${\check
S}={\hat S}(1+\hat\tau_z)/2+\hat{\widetilde S}(1-\hat\tau_z)/2$, where
$\hat{\widetilde S}({\bm p}_{f\|})={\hat S}^{tr}(-{\bm p}_{f\|})$ and
\begin{equation}
\hat S=
\left(\begin{array}{cc}
r_{\uparrow \uparrow} & r_{\uparrow \downarrow} \\
r_{\downarrow \uparrow} & r_{\downarrow \downarrow}
\end{array}\right)
= \dfrac{1}{2} \left[ r_\uparrow + r_\downarrow +
(r_\uparrow - r_\downarrow) \dfrac{\bm h + \bm w}{|\bm h + \bm
w|} \bm \sigma \right] \label{SmatrixFSO1} .
\end{equation}
Here $r_{\uparrow,\downarrow} = e^{i \Theta_{\uparrow,\downarrow}}$
and, assuming spatially constant $\bm h$ and $\alpha$ in the FSOL,
\begin{equation}
\Theta_{\uparrow,\downarrow} = \pi + 2 {\rm arctan}
\left[ \dfrac{|p_{fx}|}{p_{fx \uparrow,\downarrow}}\tan (p_{fx
\uparrow,\downarrow}d) \right] - 2 |p_{fx}| d
\label{Theta_up_down_FSO1},
\end{equation}
where Fermi momenta in the normal metal ${\bm p}_f$ and in the FSOL
${\bm p}_{f\uparrow,\downarrow}$ satisfy the relation
$p^2_{fx \uparrow,\downarrow}=p^2_{fx}\pm 2m|{\bm h}+{\bm w}({\bm p}_{f\|})|$.

Making use of explicit expression for the $\check S$-matrix
\eqref{SmatrixFSO1} and following the quasiclassical approach with
Riccati amplitudes in describing the superconductor \cite{eschrig00,fogel00},
we obtain the quasiclassical Green's function. In particular, we obtain
spectra of Andreev surface states, which take the following form now
\begin{equation}
\varepsilon_{1,2} = {\rm sgn}\left[\sin \left( \dfrac{X \mp
\Phi}{2} \right)\right] \Delta\cos \left( \dfrac{X \mp \Phi}{2} \right)
\label{bsFSO1} \enspace .
\end{equation}
Here $X(\bm p_{f\|})=\frac{1}{2}(\Theta_{\uparrow}(\bm p_{f\|})+
\Theta_{\downarrow}(\bm p_{f\|})- \Theta_{\uparrow}(-\bm p_{f\|}) -
\Theta_{\downarrow}(-\bm p_{f\|}))$ and $\Phi(\bm p_{f\|})$
is defind in Eq.(\ref{Phi}), where one should use the generalized
definition for $\Theta_{\pm}({\bm p}_{f\|})$:\,
$\Theta_{\pm}({\bm p}_{f\|}) = \Theta_{\uparrow}({\pm\bm p}_{f\|})-
\Theta_{\downarrow}(\pm\bm p_{f\|})$. For a small parameter
$|{\bm h}+{\bm w}|/\varepsilon_{f}\ll 1$ the quantity $X(\bm p_{f\|})$
vanishes in the first approximation, while the definition for $\Theta_{\pm}$
reduces to that given in Eq.(\ref{Theta}).

In general, energies $\varepsilon_{1,2}(\bm p_{f\|})$ in Eq.(\ref{bsFSO1})
are situated asymmetrically with respect to the Fermi level for a
given $\bm p_{f\|}$. Since $X(\bm p_{f\|})$ and $\Phi(\bm p_{f\|})$ are odd
and even functions of $\bm p_{f\|}$ respectively, each energy branch
$\varepsilon_{1,2}(\bm p_{f\|})$ in Eq.(\ref{bsFSO1}), as well as the
Andreev spectra as a whole, is neither odd nor even with respect to the
transformation $\bm p_{f\|}\to-\bm p_{f\|}$:
$\varepsilon_{1,2}(-\bm p_{f\|})=-\varepsilon_{2,1}(\bm p_{f\|})$.
As a result of the asymmetry, the spontaneous electric current density
$j_y(x)$, flowing along the surface perpendicular to the exchange field in the
superconducting region, arises near the surface with the FSOL. The spontaneous
surface current density at the interface $x=d$ takes comparatively simple
form in the case of small spin-orbit coupling $\alpha p_f\ll
(\varepsilon_f\pm h)$:
\begin{eqnarray}
j_{y}(d)=\quad &\nonumber\\
=\dfrac{\pi e N_f\Delta}{ 2}&\left\langle v_{fy}
\left(\dfrac{\Delta}{ 2T}
\sin^2 \dfrac{\Theta_0}{ 2}\cosh^{-2}\dfrac{
\Delta \cos \Theta_0}{ 2T}-
\right.\right.\nonumber \\
&\left.\left.
-\cos\dfrac{\Theta_0}{ 2}\tanh \dfrac{
\Delta \cos \Theta_0}{ 2T} \right)
X(\bm p_{f\|}) \right\rangle_{S_f} .
\label{elcurd}
\end{eqnarray}
Here $\Theta_0$, taken for zero spin-orbit coupling, is defined as
$\Theta_0=\Theta_{+}|_{\alpha=0}=\Theta_{-}|_{\alpha=0}$. The
expression for $X(\bm p_{f\|})$ in Eq.(\ref{elcurd}) should be taken
linear in small parameters $\alpha p_f/(\varepsilon_f\pm h)$. Then
$X(\bm p_{f\|})\propto w_z=\alpha p_{fy}$ and averaging over the
Fermi surface in Eq.\eqref{elcurd} gives nonzero result for $j_{y}$,
while $j_z$ vanishes.

We notice, that an expression for the Josephson critical current in S-F-S
junctions with small momentum dependent transparencies $D({\bm p}_{f\|})$
\cite{nazarov01,bb02} can be obtained from Eq.\eqref{elcurd}
by replacing $X(\bm p_{f\|})v_{fy}\to -2D({\bm p}_{f\|})|v_{fx}|$.
This is not surprising, since both the spontaneous surface current and
the Josephson current are actually the two components of total supercurrent
carried by the same Andreev interface states, which reduce to surface states
in the tunneling limit. In the particular case $h\sim\alpha p_f\ll
\varepsilon_f$ the spontaneous surface electric current $j_y\propto \alpha
p_fh/\varepsilon_f^2$ is of the second order in a small parameter
$(h/\varepsilon_f)\sim(\alpha p_f/\varepsilon_f)$. Since these small
second-order terms are disregarded within the quasiclassical approach to
describing the FSOL, solutions of Eq.(\ref{eilenberger}) found above show
no spontaneous electric surface current, in contrast with the spin currents
in the FSOL.

{\it Acknowledgments} This work was supported by the Russian Foundation for
Basic Research under grant 02-02-16643, scientific programs of Russian
Ministry of Science and Education and Russian Academy of Sciences.
I.V.B. acknowledges the support of the Dynasty Foundation and
thanks the Forschungszentrum J\"ulich for financial
support within the framework of the Landau Program.


\end{document}